\documentclass[12pt,a4paper]{article}

\usepackage{graphics}
\usepackage{latexsym}
\usepackage{amsmath}
\usepackage{amssymb}
\usepackage{slashed}
\usepackage[usenames]{color}

\title{Measurement of the Dark Matter Velocity Dispersion with
Galaxy Stellar Masses, UV Luminosities, and Reionization}
\author{Bruce Hoeneisen}
\date{\small{
Universidad San Francisco de Quito, Quito, Ecuador \\
Email: bhoeneisen@usfq.edu.ec \\
18 August 2022}
}

\begin{document}
\maketitle

\begin{abstract}
\noindent
The root-mean-square of	non-relativistic warm dark matter particle velocities scales as
$v_{h\textrm{rms}}(a) = v_{h\textrm{rms}}(1)/a$, where $a$ is the
expansion parameter of the universe. This velocity dispersion results in
a cut-off of the power spectrum	of density fluctuations due to dark matter free-streaming.
Let $k_\textrm{fs}(t_\textrm{eq})$ be the free-streaming comoving cut-off
wavenumber at the time of equal	densities of radiation and matter.
We obtain $v_{h\textrm{rms}}(1) = 0.41^{+0.14}_{-0.12} \textrm{ km/s}$, 
and $k_\textrm{fs}(t_\textrm{eq}) = 2.0^{+0.8}_{-0.5} \textrm{ Mpc}^{-1}$,
at 68\% confidence,
from the observed distributions	of galaxy stellar masses and
rest frame ultra-violet luminosities. 
This result is consistent with reionization.
From the velocity dispersion cut-off we obtain the limits $v_{h\textrm{rms}}(1) < 0.54 \textrm{ km/s}$,
and $k_\textrm{fs}(t_\textrm{eq}) > 1.5 \textrm{ Mpc}^{-1}$.
These results are
in agreement with previous measurements based on spiral galaxy rotation 
curves, and on the formation of first galaxies and reionization.
These measured parameters determine the temperature-to-mass
ratio of warm dark matter. This	ratio happens to be in agreement
with the no freeze-in and no freeze-out	warm dark matter scenario of 
spin 0 dark matter particles decoupling early on from 
the standard model sector. 
Spin 1/2 and spin 1 dark matter are disfavored if nature has
chosen the no freeze-in and no freeze-out scenario.
An extension of the standard model of quarks and leptons,
with scalar dark matter	that couples to	the Higgs boson, that is
in agreement with all current measurements, is briefly reviewed.
Discrepancies with limits on dark matter particle mass that can 
be found in the literature,
are addressed.
\end{abstract}


\section{Introduction}

Let $v_{h\textrm{rms}}(a)$ be the root-mean-square velocity of 
non-relativistic dark matter particles.
This velocity dispersion scales with the expansion parameter $a$ of the universe as 
$v_{h\textrm{rms}}(a) = v_{h\textrm{rms}}(1)/a$ (assuming collisions, if any, do
not excite internal degrees of freedom), so
\begin{equation}
v_{h\textrm{rms}}(1) = v_{h\textrm{rms}}(a) a = 
v_{h\textrm{rms}}(a) \left[ \frac{\Omega_c \rho_\textrm{crit}}{\rho_h(a)} \right]^{1/3},
\label{vhrms1}
\end{equation}
is an adiabatic invariant. 
\footnote{Two ways to understand (\ref{vhrms1}) are:
1) Consider an expanding universe. The velocity of a free particle with respect to that comoving observer
that is momentarily at the position of the particle, is proportional to $a^{-1}$.
To obtain this result use Hubble's law with $H = a^{-1} da/dt$.
2) The adiabatic expansion of a collisional, or collisionless, noble
gas satisfies $T / \rho^{\gamma-1} =$ constant, with
$\gamma = 5/3$. Since $T \propto \left< v^2 \right>$, and $\rho \propto a^{-3}$,
$\left< v^2 \right> \propto a^{-2}$.}
$\rho_h(a)$ is the dark matter density.
(We use the standard notation in cosmology, and parameters, as in \cite{PDG2020}).
In the cold dark matter $\Lambda$CDM cosmology it is assumed that 
dark matter velocity dispersion is negligible. $v_{h\textrm{rms}}(1)$ is the single
parameter that is added to the $\Lambda$CDM model to obtain the warm dark matter
cosmology $\Lambda$WDM.
Let $P(k)$ be the power spectrum of relative density perturbations, referred to the
present time, in the $\Lambda$CDM scenario. 
$k$ is the comoving wavenumber.
Then the power spectrum of $\Lambda$WDM is
$P(k) \tau^2(k)$, where $\tau^2(k)$ is a cut-off factor due to free-streaming of
dark matter particles.
At the time $t_\textrm{eq}$ of equal radiation and matter densities, the 
free-streaming cut-off factor has the approximate form \cite{Boyanovsky}
\begin{equation}
\tau^2(k) \approx \exp{\left[-k^2/k^2_\textrm{fs}(t_\textrm{eq})\right]},
\label{tau2}
\end{equation}
where the comoving cut-off wavenumber is \cite{Boyanovsky}
\begin{equation}
k_\textrm{fs}(t_\textrm{eq})
= \frac{1.455}{\sqrt{2}} \sqrt{\frac{4 \pi G \bar{\rho}_m(1) a_\textrm{eq}}{v_{h\textrm{rms}}(1)^2}},
\label{kfs}
\end{equation}
where $\bar{\rho}_m(1) \equiv \Omega_c \rho_\textrm{crit}$ is the dark matter density at the present time.
At later times the Jeans mass decreases as $a^{-3/2}$, so non-linear regeneration of small scale
structure becomes possible, and
gives $\tau^2(k)$ a ``tail" when relative density perturbations approach unity.
The challenge is to measure $v_{h\textrm{rms}}(1)$ and $k_\textrm{fs}(t_\textrm{eq})$, and
cross-check that their relation is consistent with (\ref{kfs}).

$v_{h\textrm{rms}}(1)$ has been obtained from observed spiral galaxy rotation curves
and equation (\ref{vhrms1}) \cite{part1} \cite{adiabatic_invariant} \cite{wdm_measurements_and_limits}.
$k_\textrm{fs}(t_\textrm{eq})$ has been obtained from observed
galaxy stellar mass distributions \cite{fermion_or_boson},
and from the redshift $z$ of first galaxies and reionization \cite{first_galaxies}.
In the present study we measure $k_\textrm{fs}(t_\textrm{eq})$ with both
galaxy stellar mass distributions, and galaxy rest frame ultra-violet (UV) luminosity distributions.
We also study reionization.
These measurements are compared with the predictions of the no freeze-in and no freeze-out warm dark matter
scenario as developed in \cite{wdm_measurements_and_limits}. 
Finally, an extension of the standard model of quarks and leptons 
that satisfies all current experimental constraints is briefly reviewed.

\section{Measurement of $k_\textrm{fs}(t_\textrm{eq})$}
\label{measurement}

\begin{figure}
\begin{center}
\scalebox{0.335}
{\includegraphics{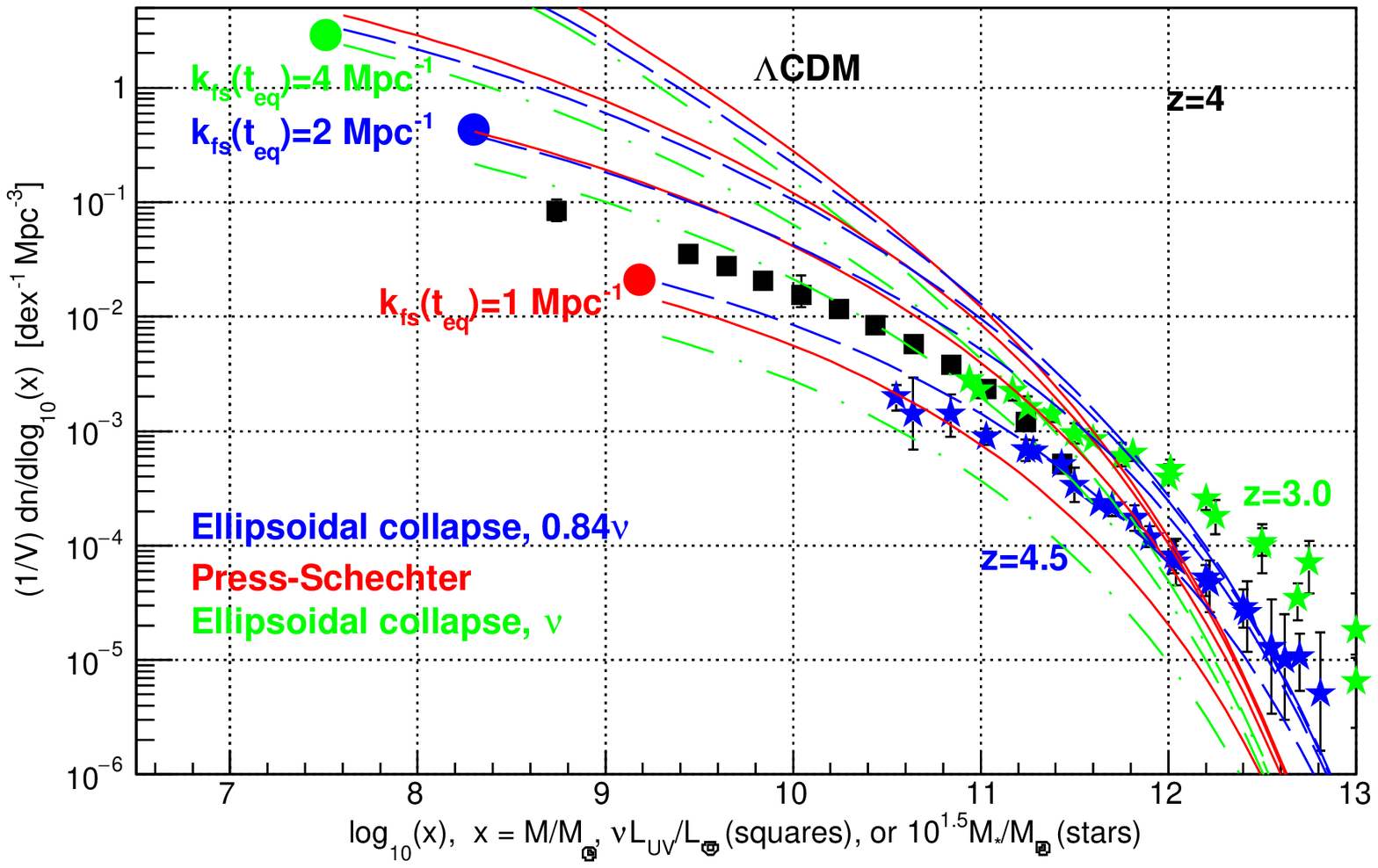}}
\scalebox{0.335}
{\includegraphics{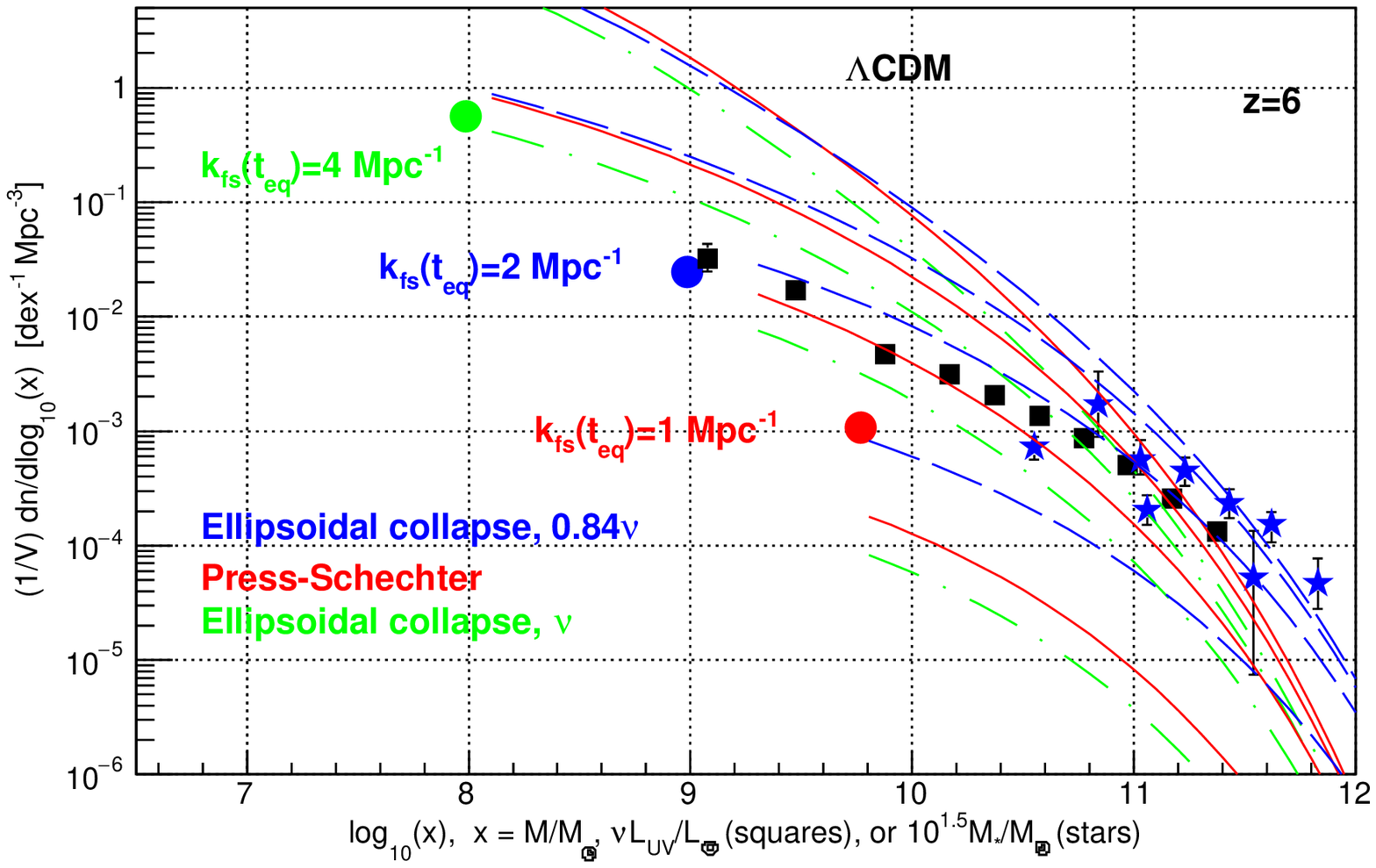}}
\scalebox{0.335}
{\includegraphics{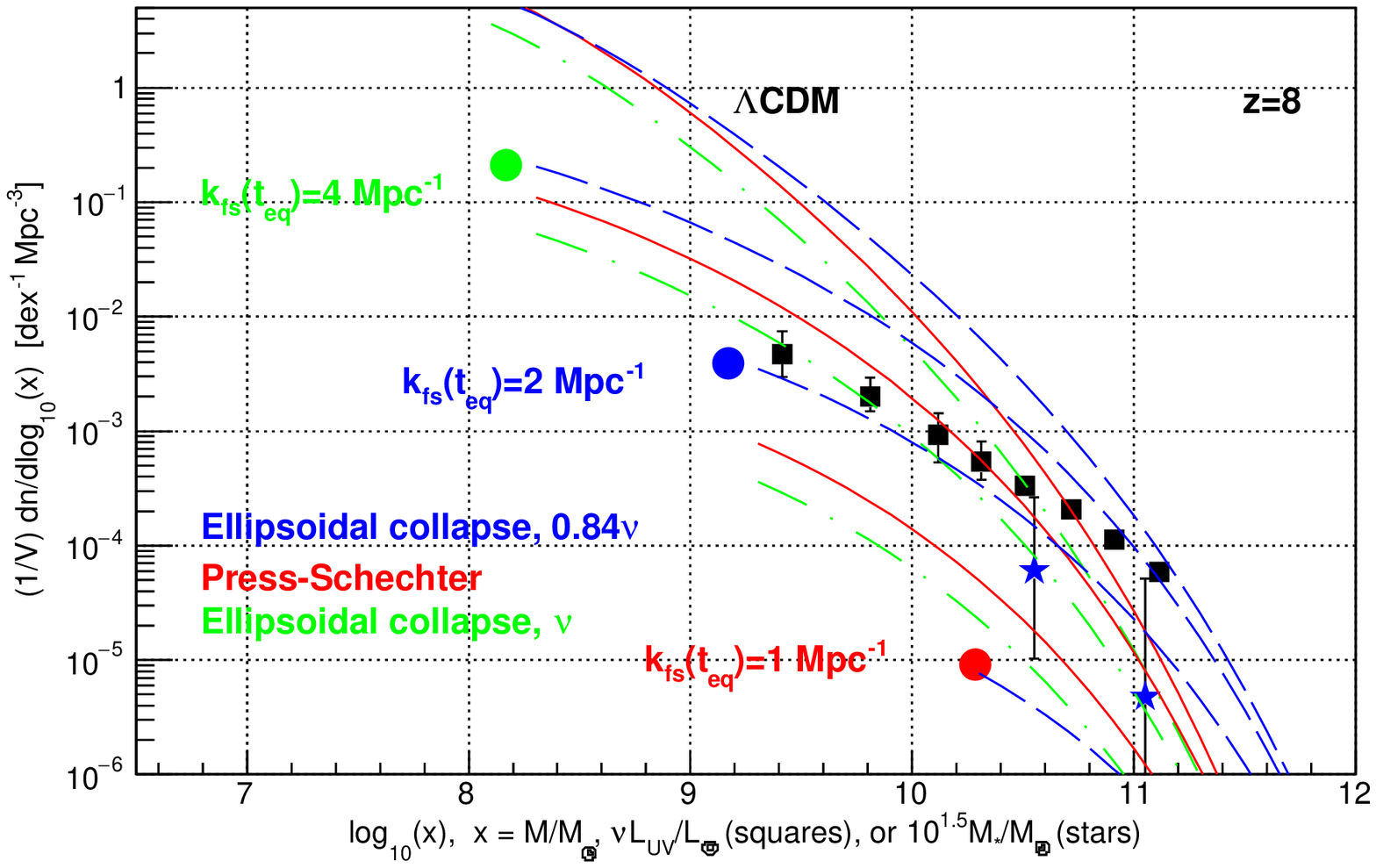}}
\scalebox{0.335}
{\includegraphics{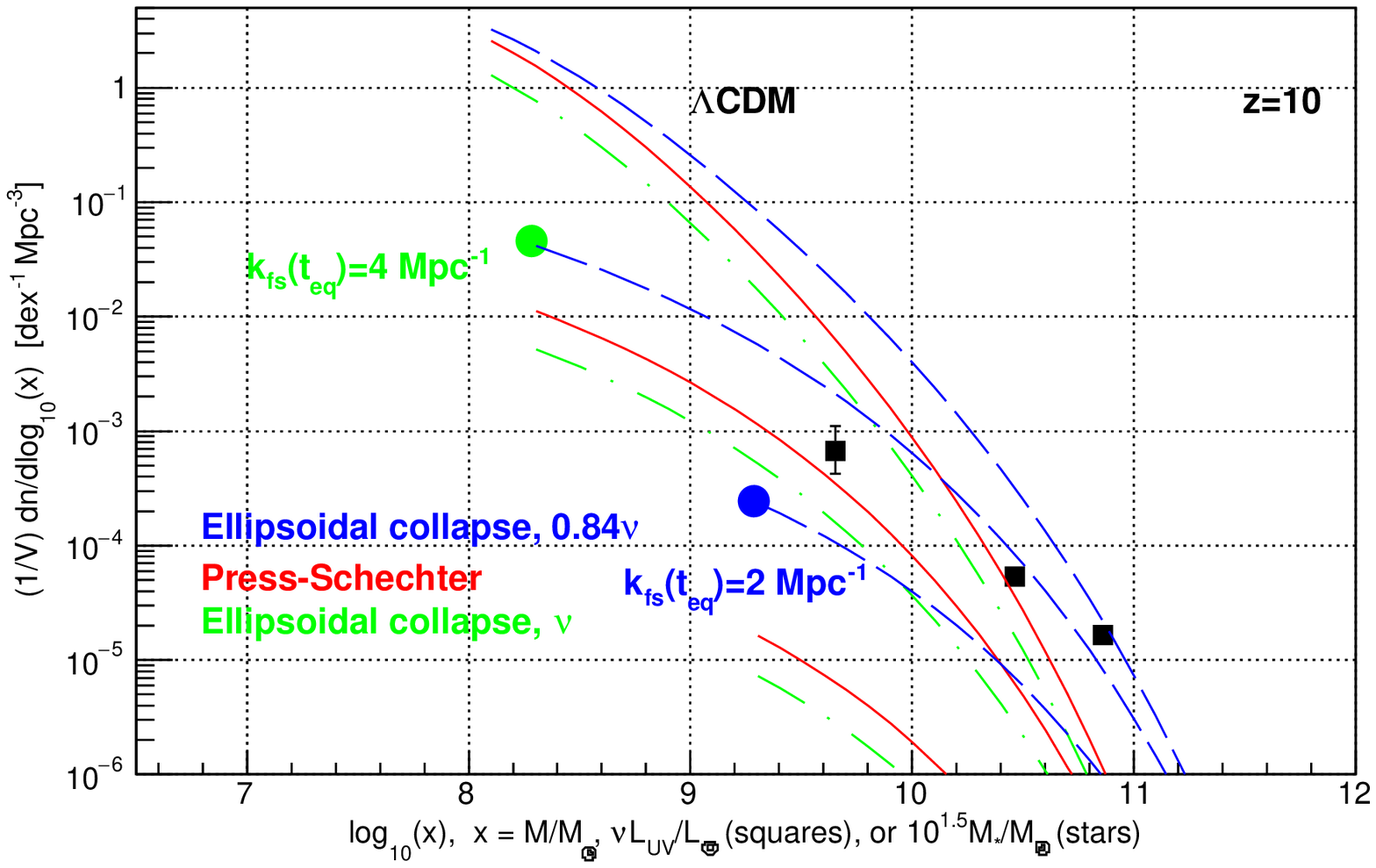}}
\caption{\small{Shown are distributions of $x$,	where $x$ is the
observed galaxy stellar mass $M_*/M_\odot$ times $10^{1.5}$ (stars), 
or the observed galaxy rest frame
ultra-violet luminosity $\nu L_\textrm{UV}/L_\odot$ (squares), or the
predicted linear total (dark matter plus baryon) mass $M/M_\odot$ (lines),
at redshifts 4, 6, 8, and 10.
The predictions correspond to $k_\textrm{fs}(t_\textrm{eq}) = 1, 2, 4$ and $1000$ Mpc$^{-1}$.
The $M_*/M_\odot$ data in the $z = 4$ panel correspond to
$z=3$ (green stars) and $z=4.5$ (blue stars).
The data sources and predictions are described in the main text.
The round red, blue and green dots indicate the velocity dispersion cut-offs of the 
predictions \cite{first_galaxies} at 
$k_\textrm{fs}(t_\textrm{eq}) = 1, 2$ and 4 Mpc$^{-1}$, respectively.
}}
\label{lintailsharpk}
\end{center}
\end{figure}

\begin{figure}
\begin{center}
\scalebox{0.335}
{\includegraphics{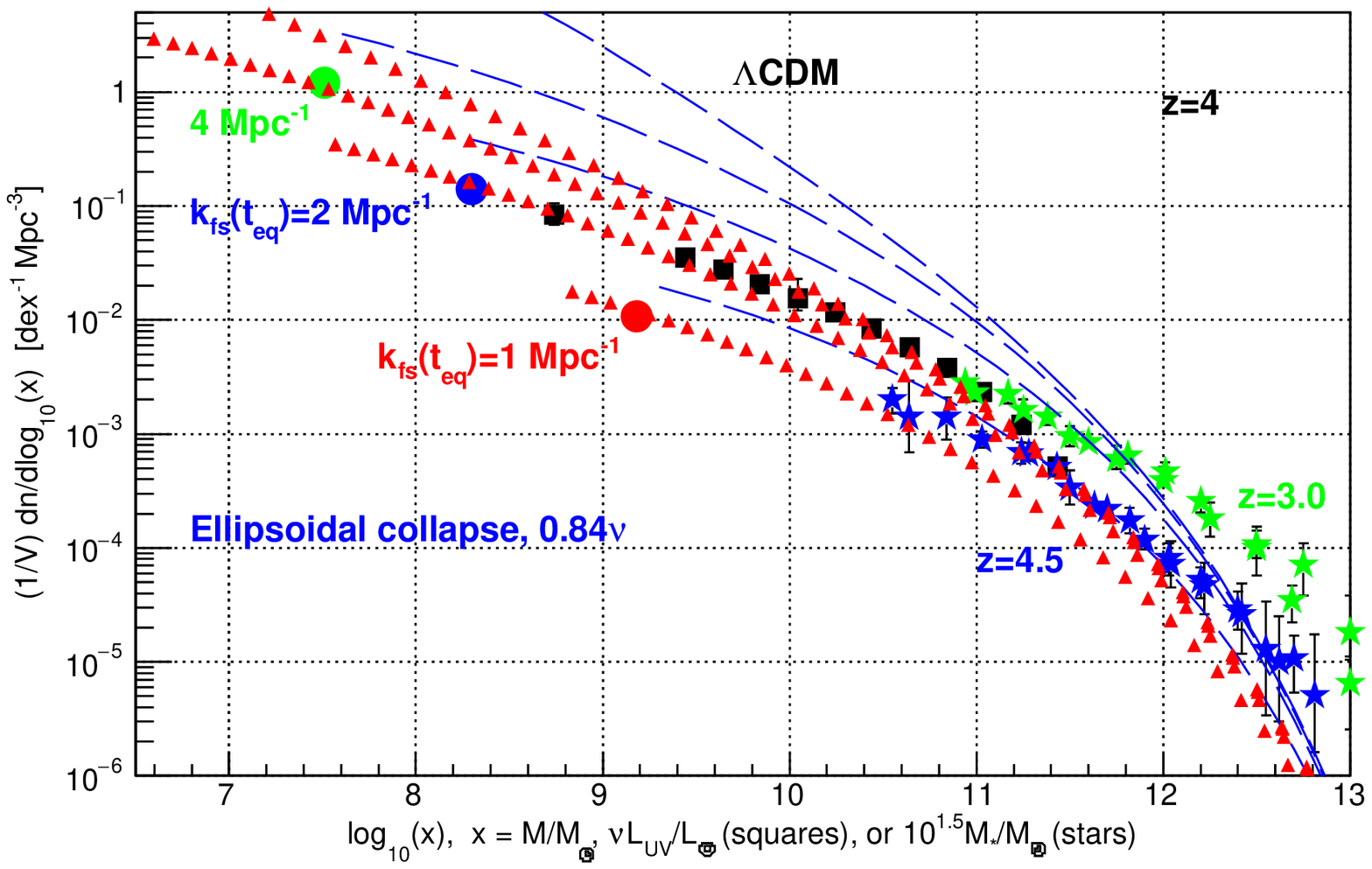}}
\scalebox{0.335}
{\includegraphics{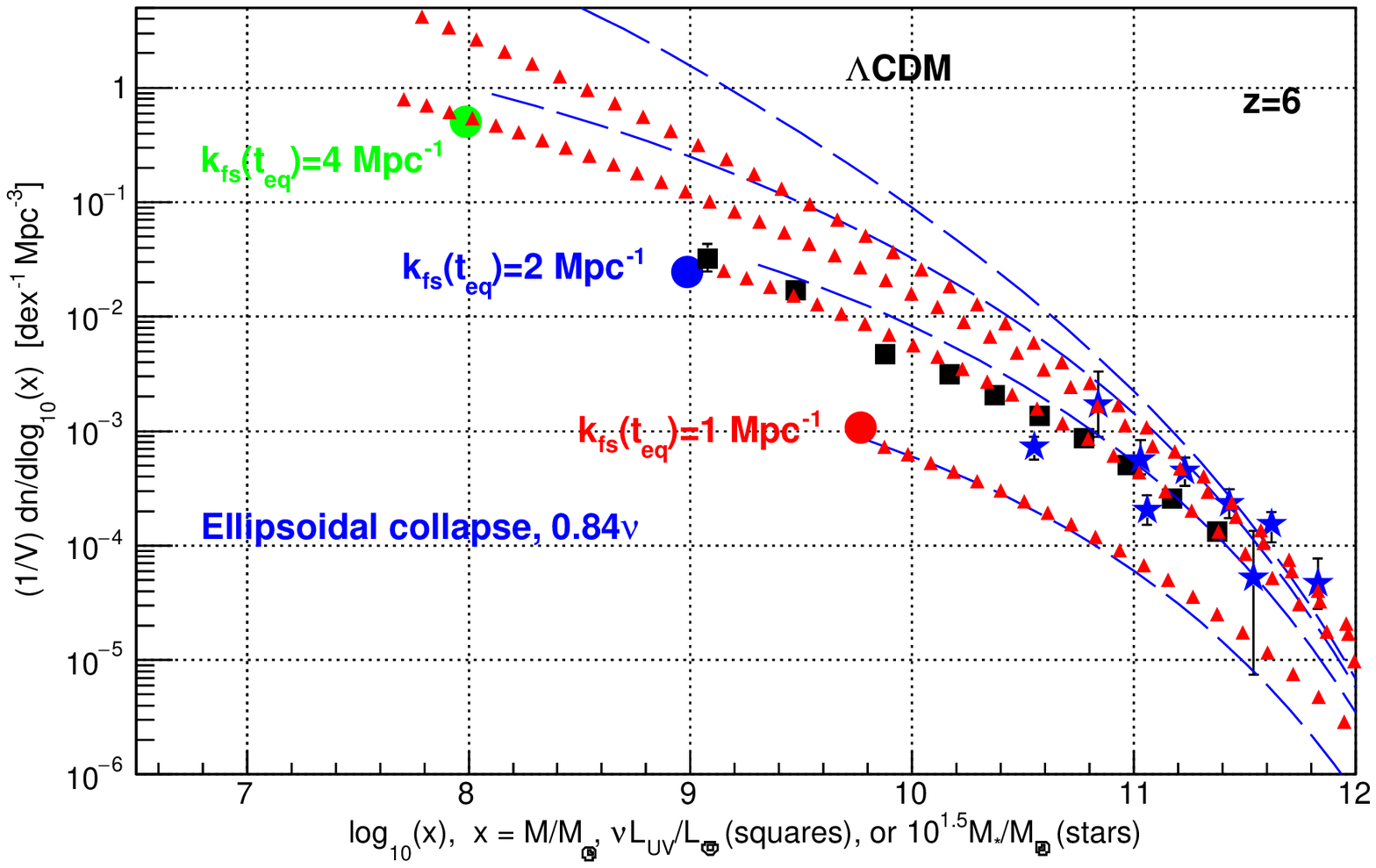}}
\scalebox{0.335}
{\includegraphics{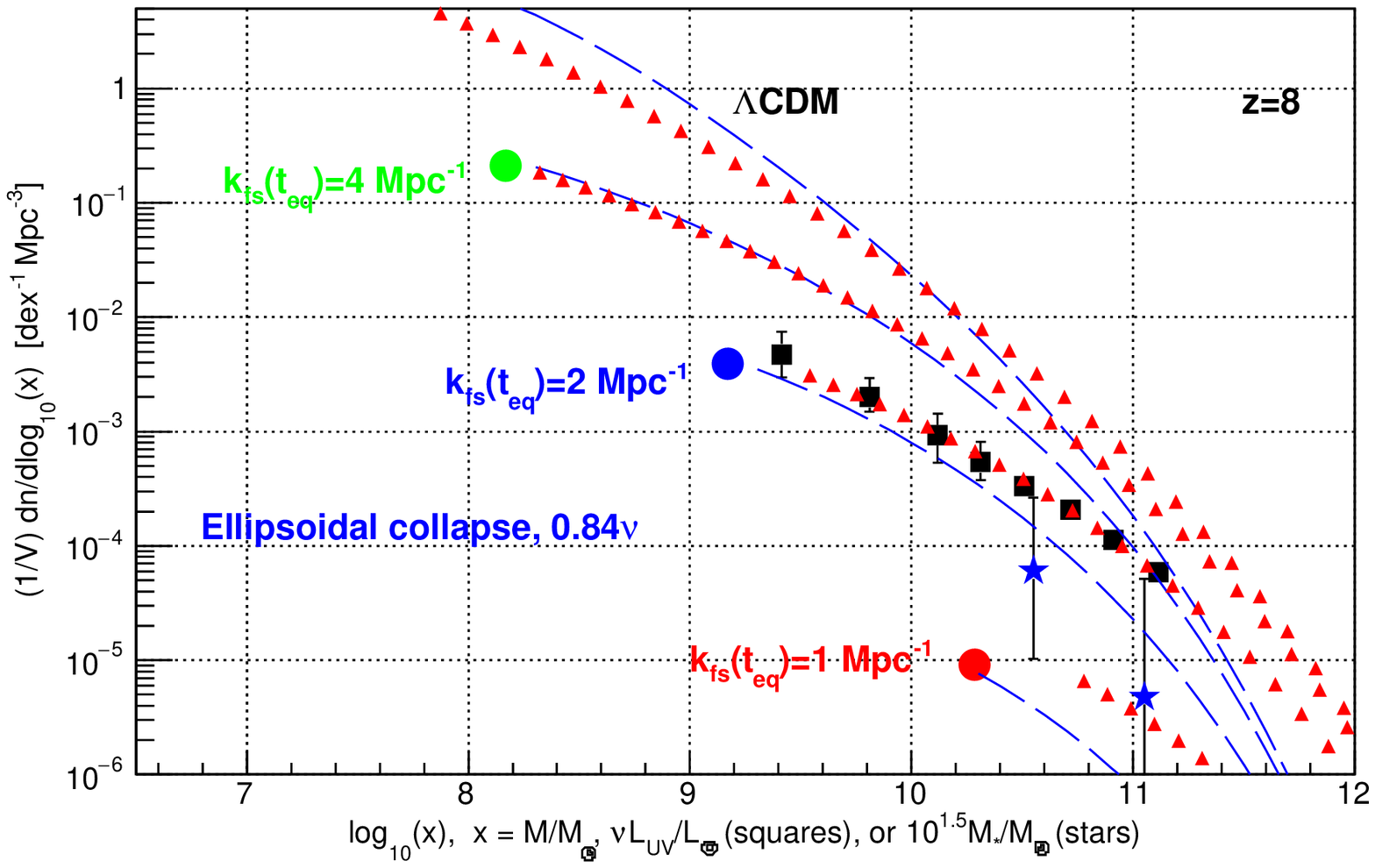}}
\scalebox{0.335}
{\includegraphics{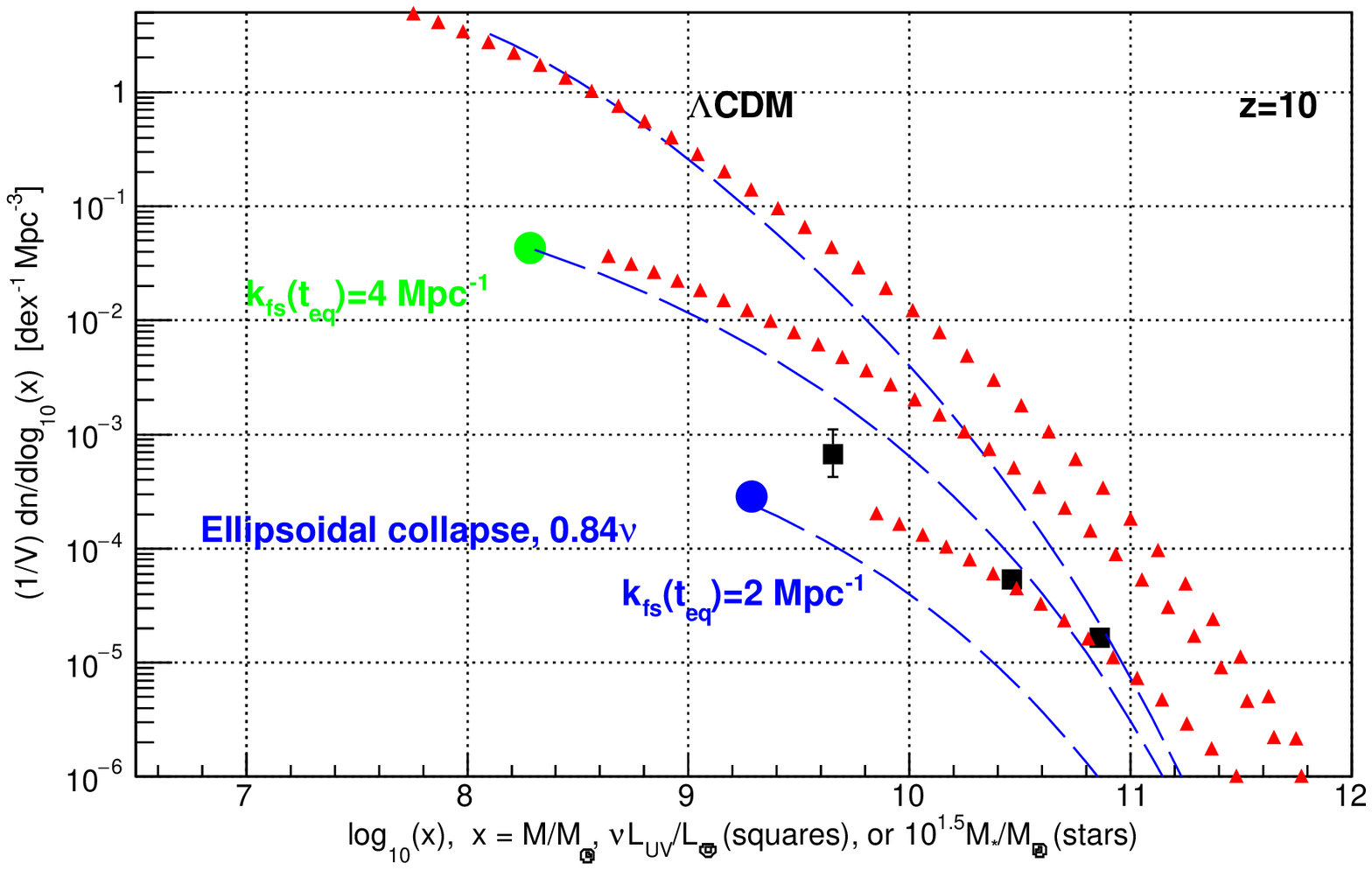}}
\caption{\small{Shown are distributions of $x$, where $x$ is the
observed galaxy stellar mass $M_*/M_\odot$ times $10^{1.5}$ (stars), or the observed galaxy rest frame
ultra-violet luminosity $\nu L_\textrm{UV}/L_\odot$ (squares), or the
predicted linear total (dark matter plus baryon) mass $M/M_\odot$ (dashed line),
or the predicted $\nu L_\textrm{UV}/L_\odot$ (triangles)
at redshifts 4, 6, 8, and 10.
The predictions correspond to $k_\textrm{fs}(t_\textrm{eq}) = 1, 2, 4$ and $1000$  Mpc$^{-1}$.
The $M_*/M_\odot$ data in the $z = 4$ panel correspond to
$z=3$ (green stars) and $z=4.5$ (blue stars).
The data sources and predictions are described in the main text.
The round red, blue and green dots indicate the velocity dispersion cut-offs of the 
predictions \cite{first_galaxies} at
$k_\textrm{fs}(t_\textrm{eq}) = 1, 2$ and 4 Mpc$^{-1}$, respectively.
}}
\label{lintailsharpk_LUV}
\end{center}
\end{figure}

Observed distributions of galaxy stellar masses, and rest-frame UV luminosities,
are compared with predictions for $k_\textrm{fs}(t_\textrm{eq}) = 1, 2, 4$, and 1000 Mpc$^{-1}$
in Figure \ref{lintailsharpk} and Figure \ref{lintailsharpk_LUV}.
The data on galaxy stellar masses $M_*$ are obtained from the compilation in \cite{Lapi_SMF},
with original measurements described in \cite{So} \cite{Gr} \cite{Da}.
The data on the rest frame UV luminosities $\nu L_\textrm{UV}$ are
obtained from the compilation in \cite{Lapi2} of
measurements with the Hubble Space Telescope \cite{Bouwens} \cite{Bouwens2021}, see also \cite{McLeod}.
The UV luminosities have been corrected for dust extinction as described in \cite{Lapi2} \cite{Bouwens_27}.
$\nu$ is the frequency corresponding to the wavelength $1550$\AA, and $L_\textrm{UV}$ is
the UV luminosity in units $\left[ \textrm{erg } s^{-1} \textrm{Hz}^{-1} \right]$.
In Figure \ref{lintailsharpk} are presented
the observed distributions of stellar masses $M_*$ and UV luminosities $\nu L_\textrm{UV}$, and
the Press-Schechter \cite{Press-Schechter} predicted distributions
of the linear total (dark matter plus baryon) mass $M$, 
and its Sheth-Tormen ellipsoidal collapse extensions with parameter
$\nu \equiv 1.686/\sigma$ (not to be confused with the frequency above) and $0.84 \nu$
\cite{Sheth_Tormen} \cite{Sheth_Mo_Tormen}. 
Our default prediction uses $0.84 \nu$.
In Figure \ref{lintailsharpk_LUV} we add a comparison with the predicted UV luminosity distributions.
The adimensional observables in the figures are $10^{1.5} M_*/M_\odot$,
$\nu L_\textrm{UV}/L_\odot$, and $M/M_\odot$, where $M_\odot = 1.988 \times 10^{30}$ kg is the solar mass,
and $L_\odot = 3.8 \times 10^{33} \textrm{erg } s^{-1}$ is the bolometric solar luminosity.
The measured UV AB-magnitudes are converted to luminosity as follows:
$M_\textrm{UV} \approx 5.9 - 2.5 \log_{10}{(\nu L_\textrm{UV}/L_\odot)}$ \cite{Lapi2}.

The Press-Schechter prediction depends on the variance of
the relative density perturbation $\delta(\textbf{x}) \equiv (\rho(\textbf{x}) - \bar{\rho})/\bar{\rho}$
on the linear total (dark matter plus baryon) mass scale $M$, at redshift $z$: \cite{Weinberg}
\begin{equation}
\sigma^2(M, z, k_\textrm{fs}) = \frac{f^2}{(2 \pi)^3 (1 + z)^2} \int_0^\infty 4 \pi k^2 dk P(k) \tau^2(k) W^2(k),
\end{equation}
and so depends on the assumed free-streaming cut-off factor $\tau^2(k)$, and on the
window function	$W(k)$ that defines the linear mass scale $M$.
We consider two window functions: 
the Gaussian window function
\begin{equation}
W(k) = \exp{\left( -\frac{k^2}{2 k_0^2} \right)}, \qquad
M = \frac{4}{3} \pi \left( \frac{1.555}{k_0} \right)^3 \bar{\rho}_m,
\end{equation}
and, in section \ref{regeneration},
the sharp-$k$ window function
$W(k) = 1$ for $k \le k_0$, $W(k) = 0$ for $k > k_0$, and
\begin{equation}
M_h = \frac{4}{3} \pi \left( \frac{c}{k_0} \right) ^3 \bar{\rho}_h.
\label{c}
\end{equation}
Figure \ref{lintailsharpk} and Figure \ref{lintailsharpk_LUV} are obtained with
the free-streaming cut-off function with a non-linear regenerated ``tail" \cite{wdm_measurements_and_limits}
\begin{eqnarray}
\tau^2(k) & = & \exp{\left( -\frac{k^2}{k^2_\textrm{fs}(t_\textrm{eq})} \right)} \qquad
\textrm{ if } k < k_\textrm{fs}(t_\textrm{eq}), \nonumber \\
& = & \exp{\left( -\frac{k^n}{k^n_\textrm{fs}(t_\textrm{eq})} \right)}\qquad
\textrm{   if } k \ge k_\textrm{fs}(t_\textrm{eq}),
\label{tail}
\end{eqnarray}
with $n = 1$,
and the Gaussian window function.
The parameter $n$ allows a study of the effect of the non-linear regenerated
tail on the measurement.
As we shall see later on in section \ref{regeneration},
the results are insensitive to $n$ in the range 0.2 to 1.1,
and in this range of $n$, the Gaussian and sharp-$k$ window functions
obtain approximately the same results.
The $\Lambda$WDM power spectrum $P(k) \tau^2(k)$ is normalized, 
for each $k_\textrm{fs}(t_\textrm{eq})$, so that
$\sigma_8 = 0.811$ with a top-hat window function of
radius $r = 8/h \textrm{ Mpc} = 8/0.674 \textrm{ Mpc}$ \cite{PDG2020}.

Let us comment on masses.
The linear perturbation	mass scale $M$ in the Press-Schechter formalism	is well
defined, since the linear perturbation dimensions scale as $a$, and the density
scales as $a^{-3}$, so
$M$ is independent of the expansion parameter $a$.
However, in the	warm dark matter scenario, the dark matter halo	mass is	ill defined:
the halo radius	grows with a constant velocity, and the halo
mass grows linearly with time indefinitely \cite{galaxy_formation}.
The relation between the linear perturbation mass $M$ and the final galaxy stellar mass $M_*$
is non-trivial:	not only must the galaxy halo form, but	the baryons must loose energy
by radiation until the baryon density $\rho_b(r)$ decreases faster than	$r^{-3}$ at large radius $r$,
so $M_*$ becomes finite and well defined \cite{first_galaxies}.
However, the comparison	between	predictions and	observations in	Figure \ref{lintailsharpk}
and Figure \ref{lintailsharpk_LUV} offers a useful empirical relation between the observables:
\begin{equation}
\frac{\nu L_\textrm{UV}}{L_\odot} \equiv a \frac{M_*}{M_\odot} \equiv b \frac{M}{M_\odot},
\label{M_LUV_Mstar}
\end{equation}
where $a \approx 10^{1.5}$ (as in \cite{fermion_or_boson}), and $b \approx 1$, independently of $M$ or $z$.
(A more detailed analysis could take $a$ and $b$ to be functions of $M$ and $z$, e.g.
$a \approx 10^2$ at $z = 8$, see Figure \ref{lintailsharpk}.)
The factor $1/a \approx 10^{-1.5}$ is equal to $0.2 \Omega_b/(\Omega_c + \Omega_b)$, indicating that approximately $20\%$ 
of the original baryons in the linear density perturbation that forms the galaxy, ends up
in the galaxy stars.

How is the UV luminosity predicted?
The Press-Schechter relation, or its Sheth-Tormen ellipsoidal collapse extensions,
obtain the numbers of collapsed halos in bins of $\log_{10}{M}$ and $z$.
From (\ref{M_LUV_Mstar}) we obtain $\log_{10}{M_*}$. This allows the calculation
of the star formation rate (SFR). Finally, the rest frame UV luminosity	
per galaxy is obtained from
\begin{equation}
\frac{\nu L_\textrm{UV}}{L_\odot} = 10^{9.61} \times \left( \frac{\textrm{SFR per halo}}{[M_\odot \textrm{yr}^{-1}]} \right),
\label{LUV}
\end{equation}
as in \cite{Madau}.
In Figure \ref{lintailsharpk_LUV} note the excellent agreement of $\nu L_\textrm{UV} / L_\odot$
data and predictions for all $z$ and $M$.

Now a word on the \textit{velocity dispersion} cut-off. 
If dark matter is warm, the formation of
galaxies has two cut-offs: the \textit{free-streaming} cut-off due to the
free-streaming cut-off factor $\tau^2(k)$ in the power spectrum of density perturbations, 
and the \textit{velocity dispersion} cut-off \cite{first_galaxies}. 
In the $\Lambda$CDM cosmology, when a spherically symmetric relative density perturbation
$(\rho - \bar{\rho})/\bar{\rho}$ reaches 1.686 in the linear approximation,
the exact solution diverges and a galaxy forms. The same is true in the $\Lambda$WDM
scenario if the linear total perturbation mass $M$ exceeds
the \textit{velocity dispersion} cut-off $M_\textrm{vd}$. The \textit{velocity dispersion} cut-off 
$M_\textrm{vd}$ is obtained by	numerical integration of hydro-dynamical equations \cite{first_galaxies}, 
with results 
summarized in Table \ref{velocity_dispersion}, and indicated, 
in Figures \ref{lintailsharpk} to \ref{n_Gaussian},  by red, blue, and
green dots for
$k_\textrm{fs}(t_\textrm{eq}) = 1, 2$, and 4 Mpc$^{-1}$, respectively.
Below the \textit{velocity dispersion} cut-off mass $M_\textrm{vd}$, the galaxy formation is delayed,
and finally no self-gravitating structure forms.
The Press-Schechter formalism includes the \textit{free-streaming} cut-off,
but not the \textit{velocity dispersion} cut-off.
Care must be taken not to apply the Press-Schechter formalism below the
\textit{velocity dispersion} cut-off, and care must be taken to include the
non-linear regenerated tail of $\tau^2(k)$.

The comparison of data and predictions in Figure \ref{lintailsharpk} favor
$k_\textrm{fs}(t_\textrm{eq}) \approx 1.5$ Mpc$^{-1}$ at $z = 4$,
increasing to $k_\textrm{fs}(t_\textrm{eq}) \approx 3.0$ Mpc$^{-1}$ at $z = 10$.
The comparison of data and predictions in Figure \ref{lintailsharpk_LUV}	is    
consistent with	$k_\textrm{fs}(t_\textrm{eq}) \approx 2$ Mpc$^{-1}$ for $z = 4, 6, 8$ and 10.
From Figures \ref{lintailsharpk} and \ref{lintailsharpk_LUV}, and studies to be
presented in section \ref{regeneration}, we obtain
\begin{equation}
v_{h\textrm{rms}}(1) = 0.41^{+0.14}_{-0.12} \textrm{ km/s}, \qquad \textrm{and} \qquad 
k_\textrm{fs}(t_\textrm{eq}) = 2.0^{+0.8}_{-0.5} \textrm{ Mpc}^{-1},
\label{vhrms_kfs}
\end{equation}
at 68\%	confidence.

\begin{table}
\begin{center}
{\small
\caption{\label{velocity_dispersion}
Shown is the velocity dispersion cut-off mass $M_\textrm{vd}$ of the linear total (dark matter plus baryon) 
mass $M$,
as a function of redshift $z$, and free-streaming comoving cut-off wavenumber
$k_\textrm{fs}(t_\textrm{eq})$.	At this	cut-off	mass $M_\textrm{vd}$, velocity dispersion
delays galaxy formation	by $\Delta z = 1$ (obtained from numerical integration of 
hydro-dynamical equations \cite{first_galaxies}).
}
\begin{tabular}{ccr|ccr}
\hline
\hline
$z$ & $k_\textrm{fs}(t_\textrm{eq})$ &	$M_\textrm{vd}$ & $z$ & $k_\textrm{fs}(t_\textrm{eq})$ & $M_\textrm{vd}$ \\
    & [Mpc$^{-1}$]                   &  [$M_\odot$]          &     & [Mpc$^{-1}$]                   &  [$M_\odot$]     \\
\hline
4   & 1               & $1.5 \times 10^9$ & 8 & 1               & $2 \times 10^{10}$ \\
4   & 1.66            & $3 \times 10^8$ & 8   & 1.66            & $4 \times 10^9$ \\
4   & 2               & $2 \times 10^8$ & 8   & 2               & $1.5 \times 10^9$ \\
4   & 4               & $3 \times 10^7$ & 8   & 4               & $1.5 \times 10^8$ \\
\hline
6   & 1               & $6 \times 10^9$ & 10  & 1               & $2 \times 10^{10}$ \\
6   & 1.66            & $2 \times 10^9$ & 10  & 1.66            & $4.5 \times 10^9$ \\
6   & 2               & $1 \times 10^9$ & 10  & 2               & $2 \times 10^9$ \\
6   & 4               & $1 \times 10^8$ & 10  & 4               & $2 \times 10^8$ \\
\hline
\hline
\end{tabular}
}
\end{center}
\end{table}

\section{Non-linear regeneration of small scale structure}
\label{regeneration}

\begin{figure}
\begin{center}
\scalebox{0.335}
{\includegraphics{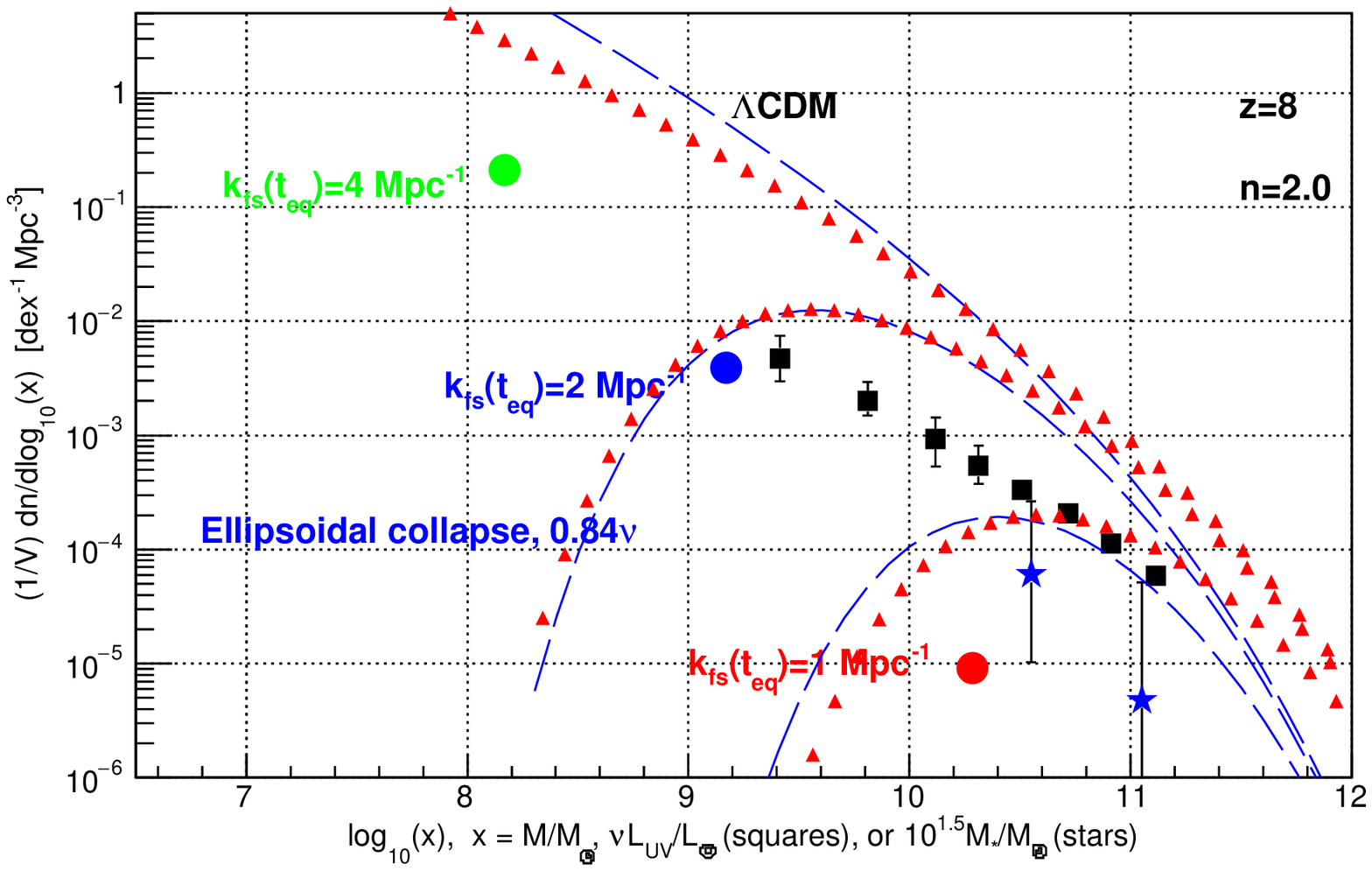}}
\scalebox{0.335}
{\includegraphics{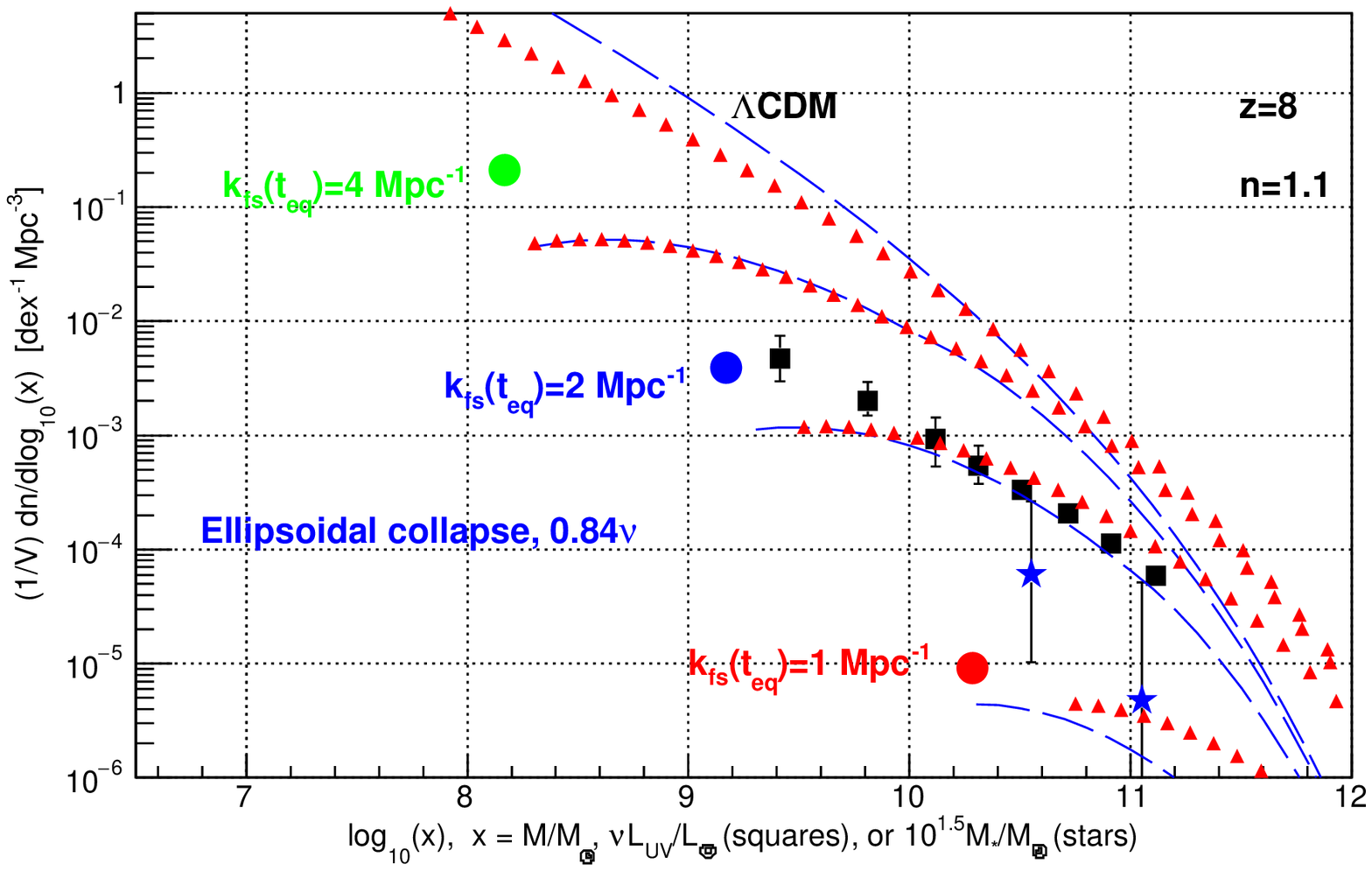}}
\scalebox{0.335}
{\includegraphics{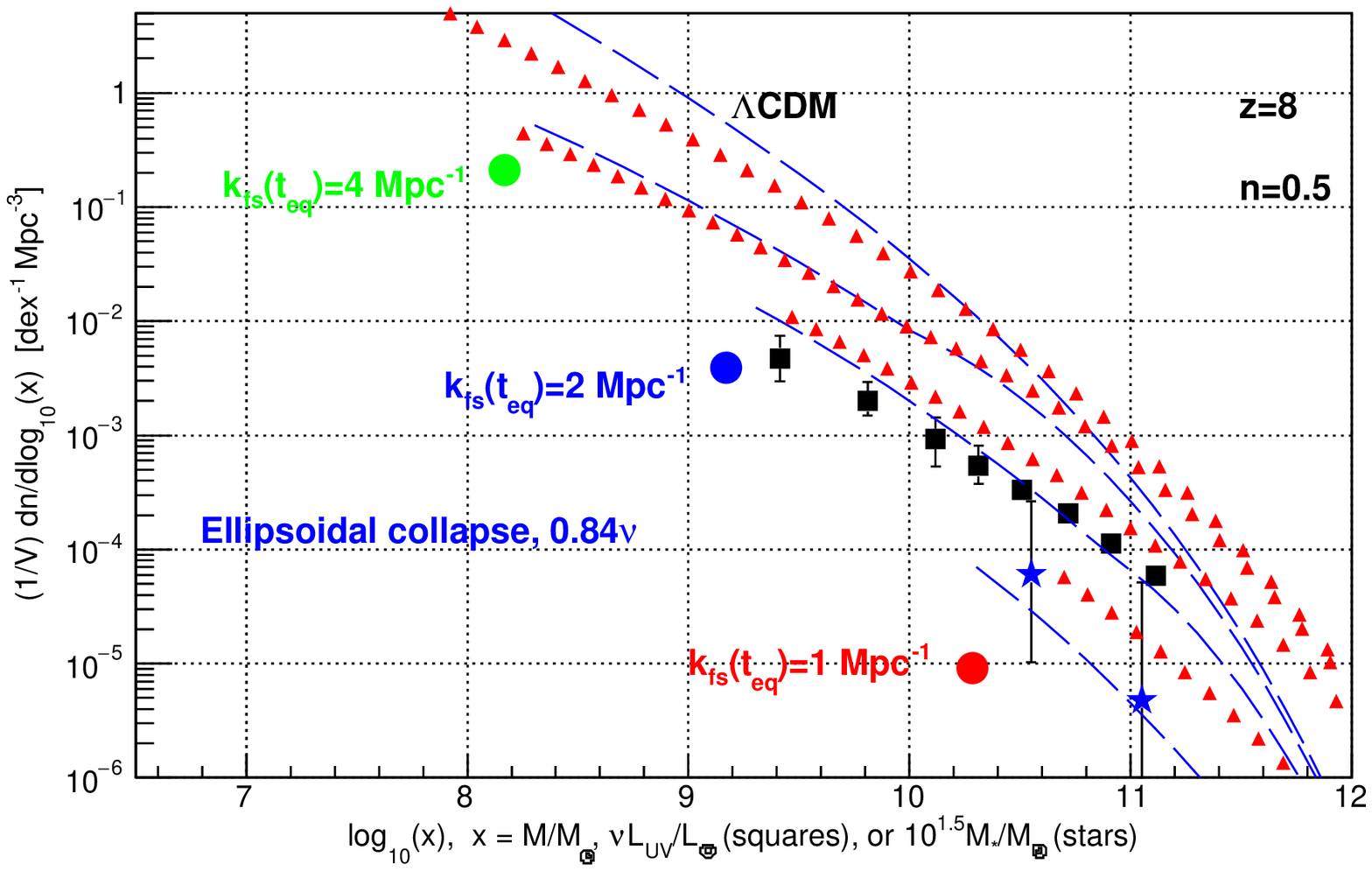}}
\scalebox{0.335}
{\includegraphics{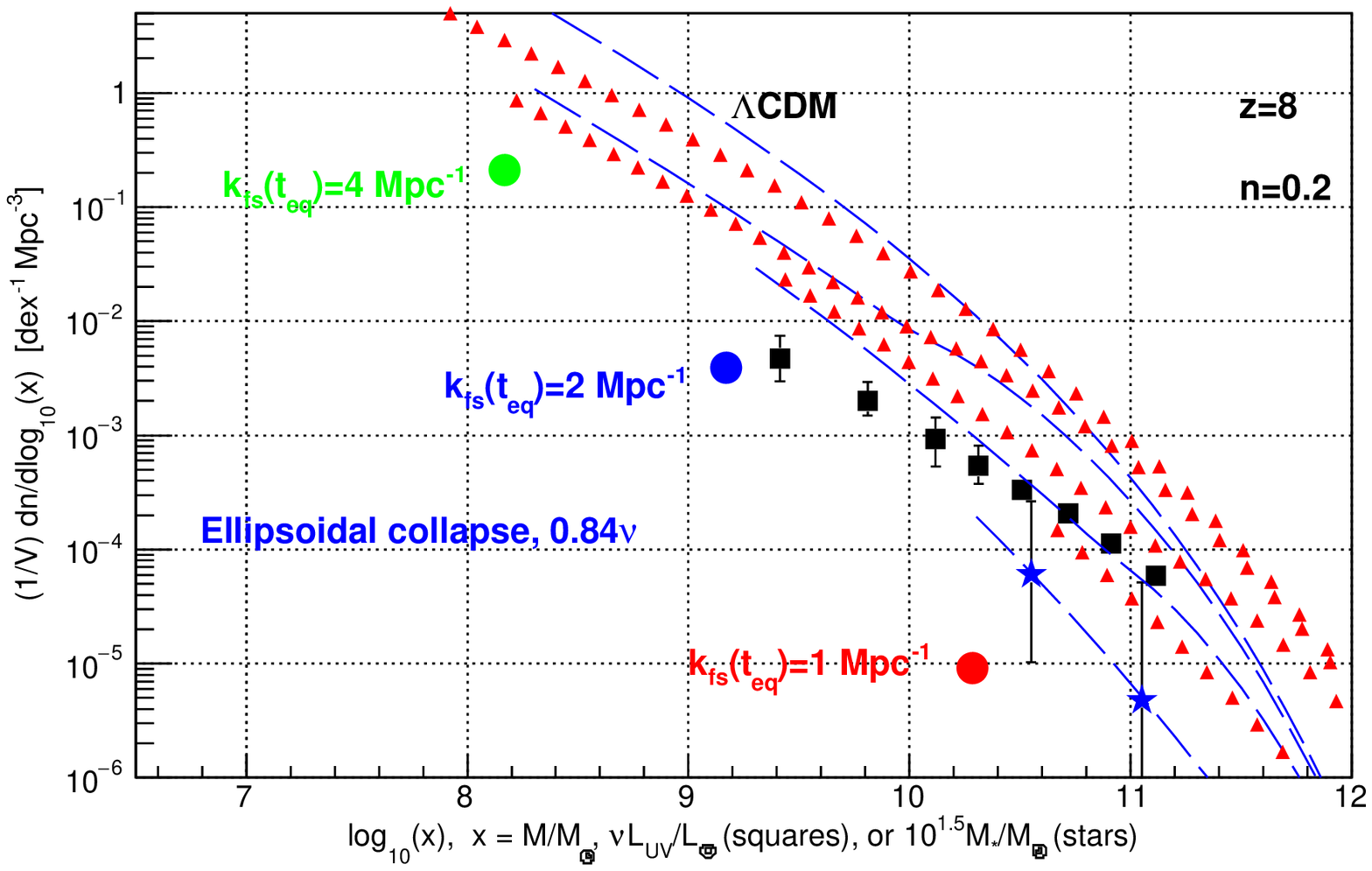}}
\caption{\small{Shown are distributions of $x$, where $x$ is the
observed galaxy stellar mass $M_*/M_\odot$ times $10^{1.5}$ (stars), or the observed galaxy rest frame
ultra-violet luminosity $\nu L_\textrm{UV}/L_\odot$ (squares), or the
predicted linear total (dark matter plus baryon) mass $M/M_\odot$ (dashed line),
or the predicted $\nu L_\textrm{UV}/L_\odot$ (triangles),
at redshift 8.
The predictions correspond to $k_\textrm{fs}(t_\textrm{eq}) = 1, 2, 4$ and $1000$ Mpc$^{-1}$.
The free-streaming cut-off factor is
$\tau^2(k)$, with a ``tail" as in (\ref{n}), with $n = 2.0, 1.1, 0.5$, or $0.2$.
The window function is sharp-$k$.
$n = 2$ corresponds to no non-linear regenerated tail. 
The round red, blue and green dots indicate the velocity dispersion cut-offs of the 
predictions \cite{first_galaxies}  at
$k_\textrm{fs}(t_\textrm{eq}) = 1, 2$ and 4 Mpc$^{-1}$, respectively.
Note that $k_\textrm{fs} = 1$ Mpc$^{-1}$ is ruled out by the velocity dispersion cut-off,
indicated by a red dot.
}}
\label{n}
\end{center}
\end{figure}

\begin{figure}
\begin{center}
\scalebox{0.335}
{\includegraphics{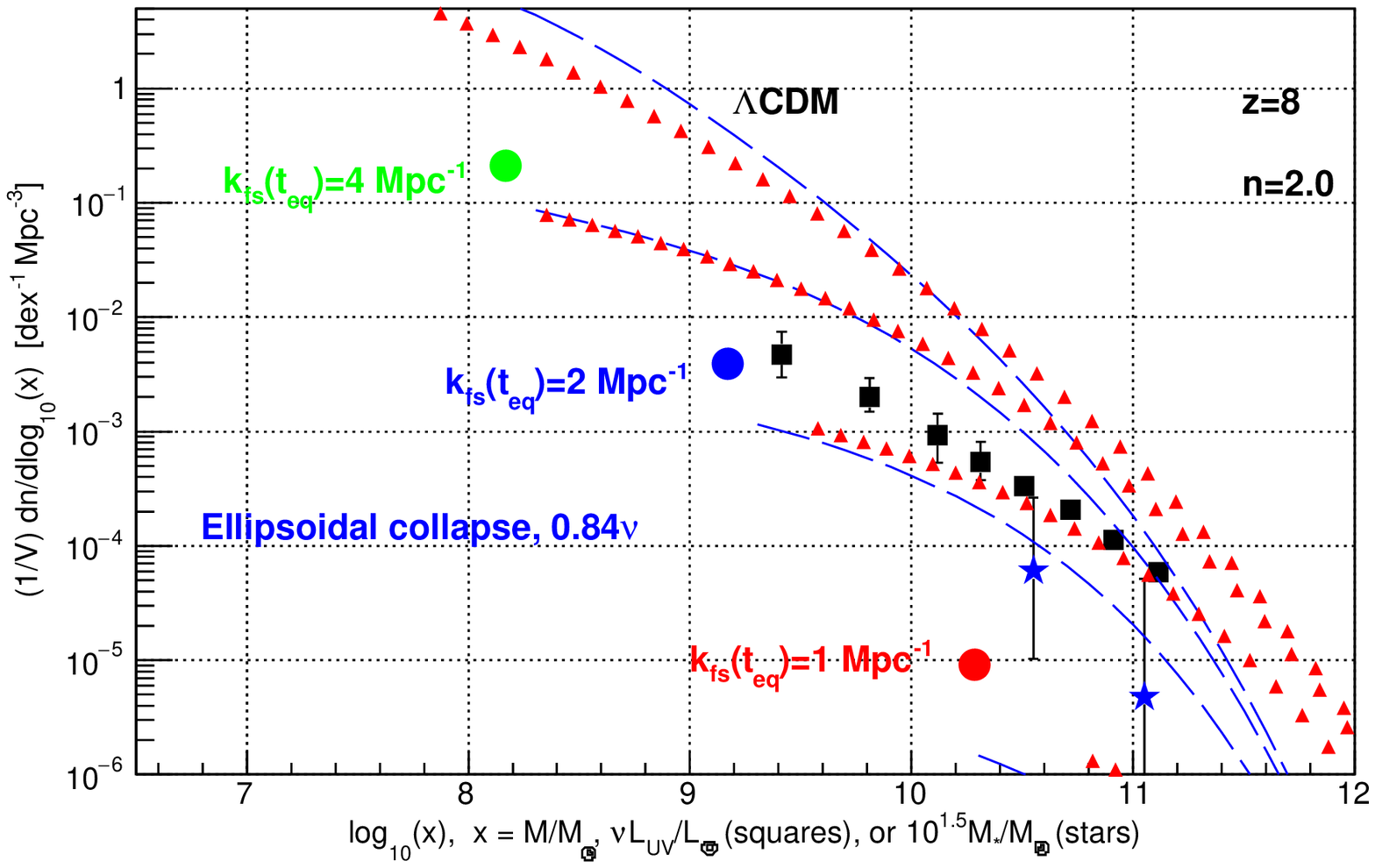}}
\scalebox{0.335}
{\includegraphics{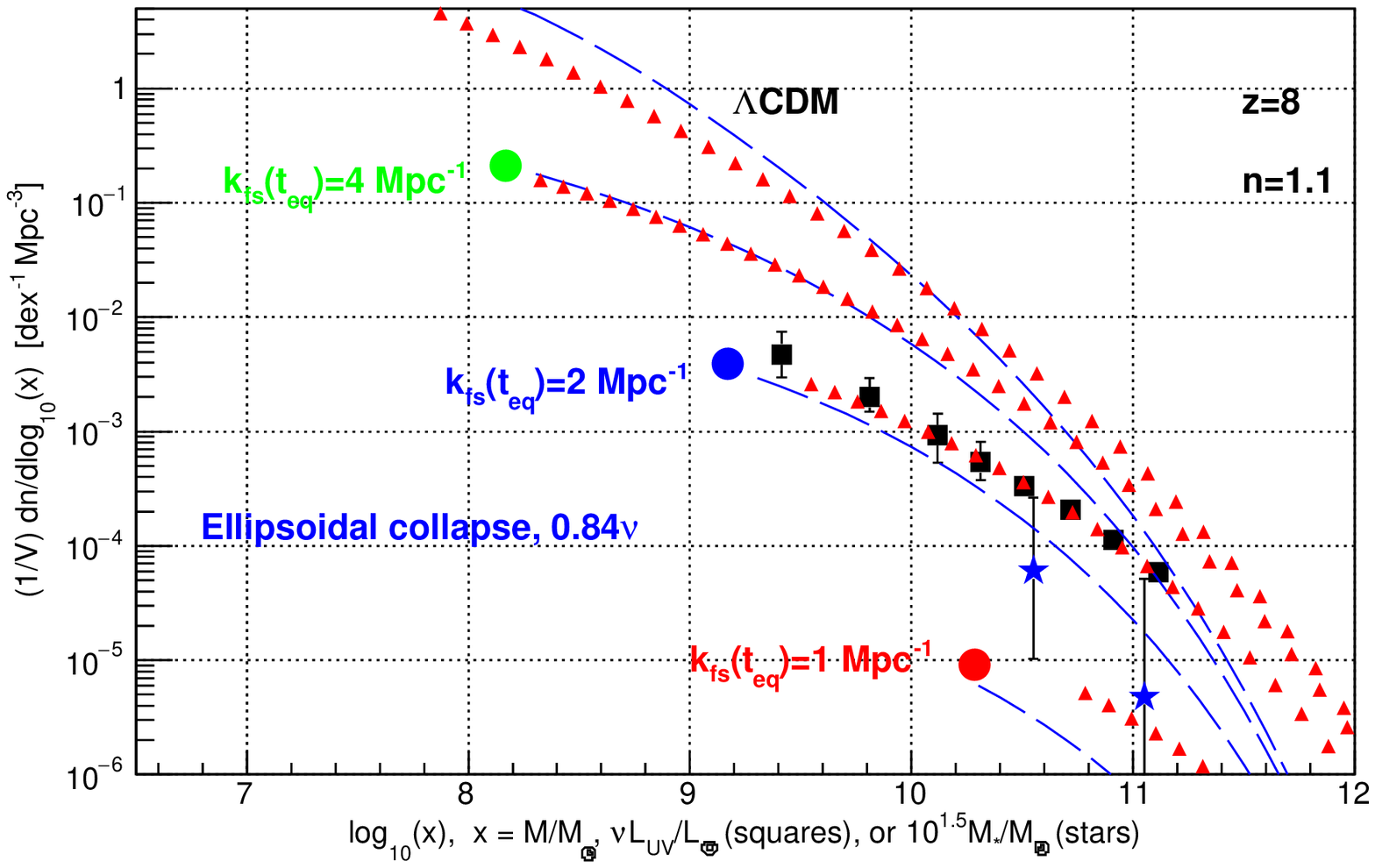}}
\scalebox{0.335}
{\includegraphics{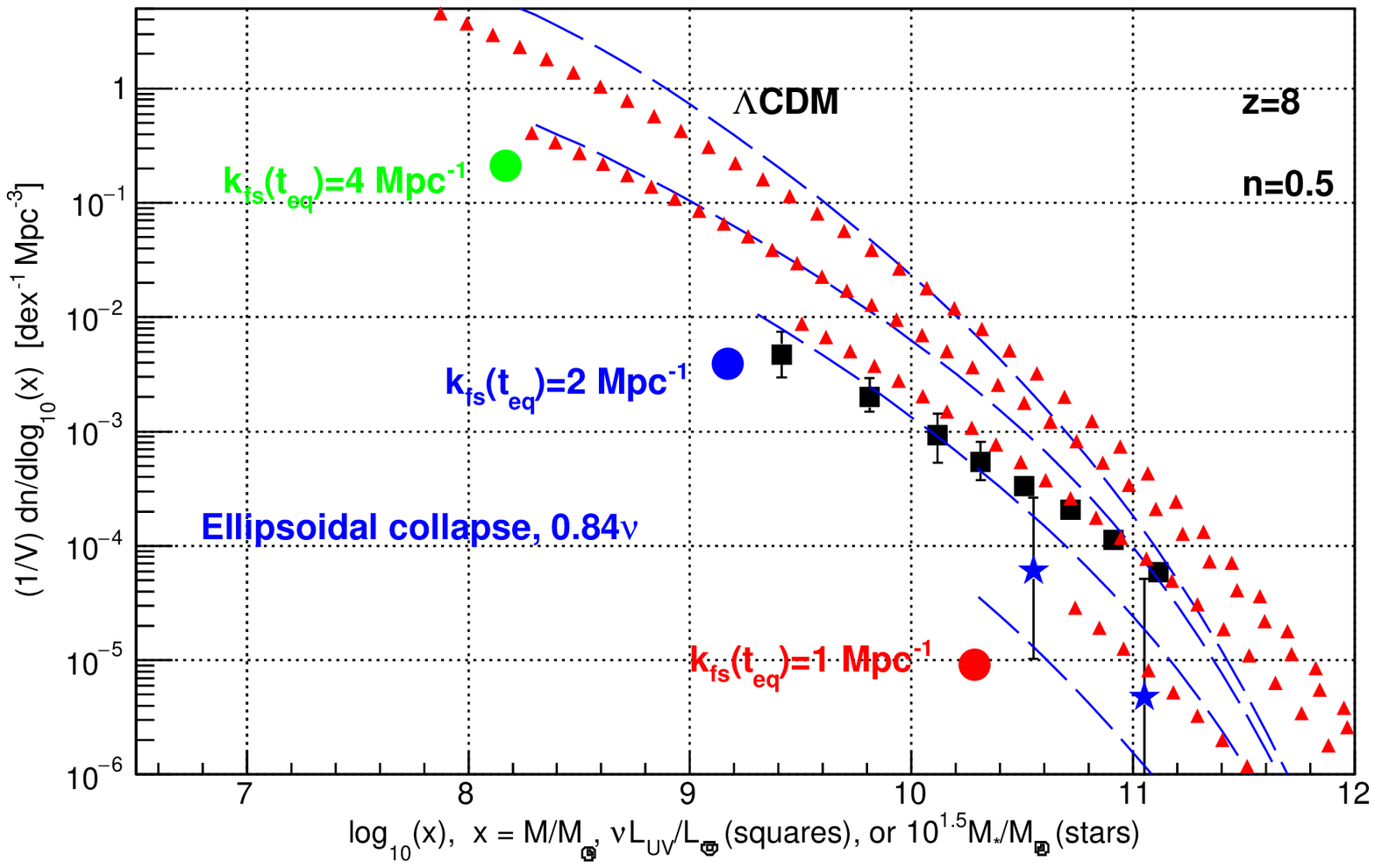}}
\scalebox{0.335}
{\includegraphics{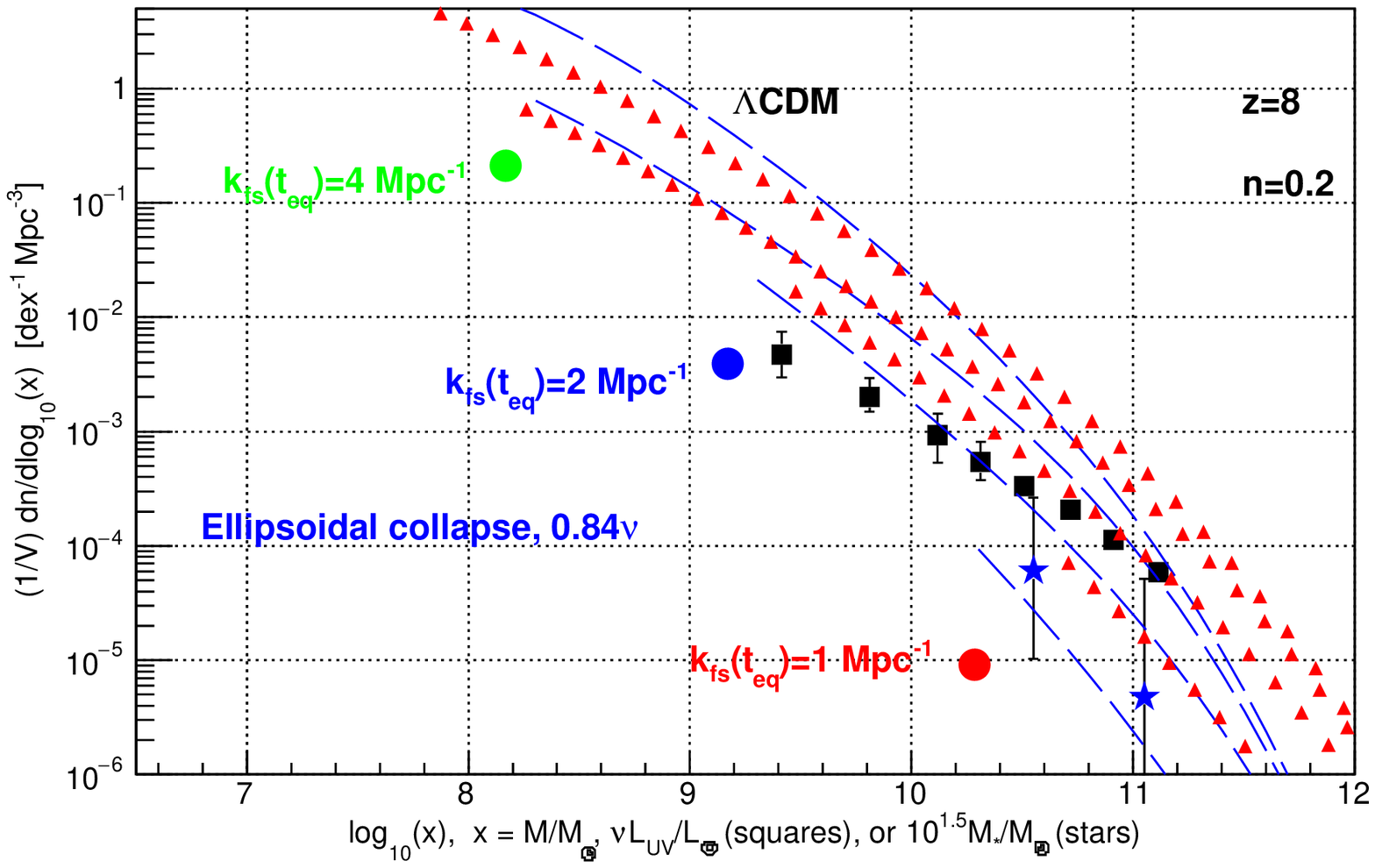}}
\caption{\small{
Same as Figure \ref{n} except that the window function is Gaussian
instead of sharp-$k$.
}}
\label{n_Gaussian}
\end{center}
\end{figure}

After equality of the densities	of radiation and matter, the 
Jeans mass decreases as	$a^{-3/2}$, allowing regeneration of
small scale structure as soon as relative density perturbations
approach unity.	The importance of this regeneration is
studied	with warm dark matter only simulations in \cite{White},
indicating that small scale structure regeneration should not be neglected. 
The uncertainty of the small scale structure regeneration contributes to
the uncertainty of the measured $k_\textrm{fs}(t_\textrm{eq})$.
To estimate this uncertainty, we perform
a data driven study by repeating Figure	\ref{lintailsharpk_LUV}
with $\tau^2(k)$ with a	regenerated tail as in (\ref{tail}), with
$n = 2.0, 1.1, 0.5$ and 0.2, and with the sharp-$k$ window function.
(Note: The sharp-$k$ window function is ill defined in $r$-space \cite{wdm_measurements_and_limits},
and has no well-defined mass $M$, so the parameter $c$ in (\ref{c}) is fixed from simulations to
$c \approx 2.7$. However, the value of $c$ does not change the measurement of
$k_\textrm{fs}(t_\textrm{eq})$, as its effect can be absorbed into the parameter $b$.
To avoid changing the value of $b = 1$, we set $c = 1.555$ \cite{wdm_measurements_and_limits}.) 
The results, for $z = 8$, are presented in
Figure \ref{n}. Agreement of observations with data is good in a wide range of $n$, i.e.
$0.5 \lesssim n \lesssim 1.1$, with
$1.6 \textrm{ Mpc}^{-1} \lesssim k_\textrm{fs}(t_\textrm{eq}) \lesssim 2.0 \textrm{ Mpc}^{-1}$.
For comparison, Figure \ref{n_Gaussian} is the same as Figure \ref{n}, except
that the Gaussian window function 
replaces the sharp-$k$ window function. The results with these two window functions
are approximately the same, except when the non-linear regenerated tail
is absent, i.e. when $n \rightarrow 2$.

The lessons learned from the studies in this section are as follows.
Nature, and simulations \cite{wdm_measurements_and_limits} \cite{White},
do indeed add a non-linear regenerated tail to the free-streaming cut-off factor $\tau^2(k)$.
With this tail, approximately the same predicted distributions are obtained
with the sharp-$k$ or Gaussian window functions, and the predictions are in
agreement with the data, so long as $0.5 \lesssim n \lesssim 1.1$.
The Gaussian window function is well behaved in both $r$-space and $k$-space.
The sharp-$k$ window function is ill behaved in $r$-space, and does not obtain
a well defined mass $M$.
Using the cut-off factor $\tau^2(k)$ without a non-linear regenerated tail, together with
the sharp-$k$ window function, leads to several published limits, of order keV, on the
warm dark matter ``thermal relic mass" that do not correspond to nature
(which does indeed regenerate a tail to $\tau^2(k)$, that should not be neglected).

\section{Reionization}

The universe is neutral from redshift $z \approx 1000$ to $z \approx 10$ when first
stars start reionizing hydrogen and helium. The bulk of reionization occurs
in the interval from $z = 8$ to $z = 6$. Thereafter hydrogen is highly ionized. 
For $z \lesssim 4$, helium becomes doubly ionized. 
The free-electrons scatter the
cosmic microwave background radiation resulting in a reionization optical depth $\tau = 0.054 \pm 0.007$
measured by the Planck collaboration \cite{PDG2020} 
(corresponding to an instantaneous reionization at $z = 7.4$).
The measured $\tau$ implies that the luminosity distributions have a cut-off \cite{Lapi2} \cite{Mason},
else the calculated $\tau$ is greater than observed. 
This is the \textit{velocity dispersion} cut-off as presented
in Table \ref{velocity_dispersion} (not the \textit{free-streaming} cut off
with a non-linear regenerated ``tail", 
and probably not a baryon physics cut-off).
From Table  \ref{tau} we obtain agreement between the Planck measurement and the
velocity dispersion cut-off, and estimate
\begin{equation}
v_{h\textrm{rms}}(1) \approx 1.2 \textrm{ to } 0.15 \textrm{ km/s}, \qquad \textrm{and} \qquad
k_\textrm{fs}(t_\textrm{eq}) \approx 0.7 \textrm{ to } 5.4 \textrm{ Mpc}^{-1}.
\label{kfs_reionization}
\end{equation}
Note that Table \ref{tau} confirms that the \textit{velocity dispersion} cut-off has physical consequences,
and implies that dark matter is indeed warm, not cold.

\begin{table}
\begin{center}
{\small
\caption{\label{tau}
At $z =	8$, for each $k_\textrm{fs}(t_\textrm{eq})$ are presented
the velocity dispersion	cut-off	$M_\textrm{vd}/M_\odot$ of the linear total
(dark matter plus baryon) mass $M/M_\odot \approx \nu L_\textrm{UV}/L_\odot$
from Table \ref{velocity_dispersion},
the corresponding cut-off AB-magnitude
$M_\textrm{UV} \approx 5.9 - 2.5 \log_{10}{(\nu L_\textrm{UV}/L_\odot)}$,
and the corresponding
reionization optical depth $\tau$ from Figure 13 of \cite{Mason}.
A somewhat lower value of $\tau$ is obtained from Figure 2 of \cite{Lapi2}.
The Planck collaboration obtains $\tau = 0.054 \pm 0.007$ \cite{PDG2020}.
}
\begin{tabular}{ccc|ccc}
\hline
\hline
$k_\textrm{fs}(t_\textrm{eq})$ & $M_\textrm{vd}/M_\odot$ & $M_\textrm{UV}$ cut-off & $\tau$ \\
\hline
1 Mpc$^{-1}$ & $2 \times 10^{10}$ & -19.9	& $0.047 \pm 0.006$ \\
2 Mpc$^{-1}$ & $1.5 \times 10^9$  & -17.0	& $0.053 \pm 0.006$ \\
4 Mpc$^{-1}$ & $1.5 \times 10^8$  & -14.5	& $0.060 \pm 0.008$ \\
\hline
\hline
\end{tabular}
}
\end{center}
\end{table}

\section{The Velocity Dispersion Cut-Off Limit}

Let us assume that the faintest UV luminosity data points in the figures
are due to the velocity dispersion cut-offs. This assumption obtains an upper
bound to $v_{h\textrm{rms}}(1)$, and a lower bound to $k_\textrm{fs}(t_\textrm{eq})$.
Allowing a factor 3 uncertainty on the velocity dispersion cut-off masses $M_\textrm{vd}$
in Table \ref{velocity_dispersion} (e.g. at what $\Delta z$ should $M_\textrm{vd}$ be defined?),
we obtain
\begin{equation}
v_{h\textrm{rms}}(1) < 0.54 \textrm{ km/s}, \qquad \textrm{and} \qquad
k_\textrm{fs}(t_\textrm{eq}) > 1.5 \textrm{ Mpc}^{-1},
\label{vel_disp}
\end{equation}
at 95\% confidence.
Note that the faintest UV luminosity data points in the figures already saturate
the reionization optical depth measured by the Planck collaboration, see Table \ref{tau}, 
and so should be near the velocity dispersion cut-offs.
A dedicated search for the cut-off of $M_\textrm{UV}$ as a function of $z$ is in order.

\section{The no Freeze-in and no Freeze-out Warm Dark Matter Scenario}
\label{no_fifo}

Table \ref{no_freeze_in_out} summarizes measurements of the velocity dispersion $v_{h\textrm{rms}}(1)$,
and the free-streaming comoving cut-off wavenumber 
$k_\textrm{fs}(t_\textrm{eq})$, as well as the no freeze-in and no freeze-out
warm dark matter scenario predictions (as developed in \cite{wdm_measurements_and_limits}).
Table \ref{no_freeze_in_out} updates Table 1 of \cite{wdm_measurements_and_limits}.
Measurements of $v_{h\textrm{rms}}(1)$ are obtained from rotation curves of 56
spiral galaxies \cite{part1} \cite{adiabatic_invariant} \cite{wdm_measurements_and_limits}.
Measurements of 
$k_\textrm{fs}(t_\textrm{eq})$ are obtained from galaxy stellar mass distributions
at $z = 4.5, 6, 7$ and 8 \cite{fermion_or_boson}. 
These measurements are indeed related by the free-streaming equation (\ref{kfs}),
within the measurement uncertainties, confirming that $k_\textrm{fs}(t_\textrm{eq})$
is due to free-streaming. 
Measurements of
$v_{h\textrm{rms}}(1)$ and $k_\textrm{fs}(t_\textrm{eq})$ are also obtained from the formation
of first galaxies and reionization \cite{first_galaxies}, and from the present
measurements (\ref{vhrms_kfs}), (\ref{kfs_reionization}) and (\ref{vel_disp}). 
These measurements determine the temperature-to-mass ratio $T(a)/m_h$ of dark matter.
This ratio happens to be in agreement with
the no freeze-in and no freeze-out
warm dark matter scenario for spin 0 dark matter particles that decouple early on
from the standard model sector, see Table \ref{no_freeze_in_out}.

We note that $k_\textrm{fs}(t_\textrm{eq}) \approx 1 \textrm{ Mpc}^{-1}$
is disfavored for several reasons:
\begin{itemize}
\item 
From the comparison of data and predictions in
Figure \ref{lintailsharpk} and Figure \ref{lintailsharpk_LUV}.
\item
First galaxies and
reionization are delayed with respect to observations \cite{first_galaxies}. 
\item
The velocity dispersion cut-off obtains the limits in (\ref{vel_disp}). For example, at $z = 8$ and 
$k_\textrm{fs}(t_\textrm{eq}) = 1 \textrm{ Mpc}^{-1}$, 
$M_{\textrm{vd}}/M_\odot = 2 \times 10^{10} \approx \nu L_\textrm{UV}/L_\odot$, see Table \ref{velocity_dispersion},
while the distribution of $\nu L_\textrm{UV}/L_\odot$ extends below this cut-off to
$2.5 \times 10^9$, see Figure \ref{lintailsharpk_LUV}.
\end{itemize}
In conclusion, if nature has chosen the no freeze-in and no freeze-out 
scenario of \cite{wdm_measurements_and_limits}, the spin 1/2 and
spin 1 dark matter alternatives are disfavored.
We note that the measurements favor scalar, i.e. spin 0, dark matter
that decouples early on from the standard model sector, e.g. scalar dark matter
coupled to the Higgs boson. In this case, the dark matter particle mass is $m_h = 150 \pm 2$ eV,
with the uncertainty mainly determined by the uncertainty of $\Omega_c h^2$ \cite{wdm_measurements_and_limits}.
Also, in this case,
\begin{equation}
v_{h\textrm{rms}}(1) = 0.49 \pm 0.01\textrm{ km/s}, \qquad \textrm{and} \qquad
k_\textrm{fs}(t_\textrm{eq}) = 1.66 \pm 0.03 \textrm{ Mpc}^{-1}.
\label{no_in_no_out}
\end{equation}

\begin{table}
\begin{center}
{\small
\caption{\label{no_freeze_in_out}Summary of measurements
of the warm dark matter velocity dispersion $v_{h\textrm{rms}}(1)$, and
the free-streaming comoving cut-off wavenumber $k_\textrm{fs}(t_\textrm{eq})$,
as well as the predictions of the no freeze-in	and no freeze-out
warm dark matter scenario \cite{wdm_measurements_and_limits}.
$a'_{h\textrm{NR}} \equiv v_{h\textrm{rms}}(1)/c$ is the expansion parameter
at which dark matter becomes non-relativistic.
After $e^+ e^-$ annihilation, while dark matter is ultra-relativistic,
$0.424 \ge T_h/T \ge \mathbf{0.344}$,
corresponding to dark matter decoupling from the standard model sector
at $T_C < T_\textrm{dec} < \mathbf{m_t}$.
* For spin 1 dark matter the predictions are model dependent \cite{adding_dm}.
** Majorana neutrino.
}
\begin{tabular}{llllll}
\hline
\hline
Observable & $v_{h\textrm{rms}}(1)$ [km/s] & $10^6 a'_{h\textrm{NR}}$ & $k_\textrm{fs}(t_\textrm{eq})$ $[\textrm{Mpc}^{-1}]$ & $m_h$ [eV] \\
\hline
Spiral galaxies \cite{part1} \cite{adiabatic_invariant} \cite{wdm_measurements_and_limits} & $0.79 \pm 0.33$ & $2.64 \pm 1.10$ & $1.03^{+0.74}_{-0.30}$ &  \\
$M_*$ distribution \cite{fermion_or_boson} & $0.91^{+0.72}_{-0.30}$ & $3.02^{+2.42}_{-0.99}$ & $0.90^{+0.44}_{-0.40}$ & \\
First galaxies \cite{first_galaxies} & $\approx 0.4$ to 0.2 & $\approx 1.4$ to 0.7 & $\approx 2$ to 4 & \\
$M_*$ and $L_\textrm{UV}$ (\ref{vhrms_kfs}) & $0.41^{+0.14}_{-0.12}$ & $1.36^{+0.45}_{-0.39}$ & $2.0^{+0.8}_{-0.5}$ & \\
Reionization (\ref{kfs_reionization}) & $\approx 1.2$ to 0.15 & $\approx 3.9$ to 0.5 & $\approx 0.7$ to 5.4 & \\
Vel. disp. cut-off (\ref{vel_disp}) & $< 0.54$ & $< 1.8$ & $> 1.5$ & \\
\hline
Fermions spin $1/2$ ** & & & & \\
No freeze-in/-out    & 1.93 to \textbf{0.83}  & 6.43 to \textbf{2.78} & 0.42 to \textbf{0.98} & 54 to \textbf{101} \\
\hline
Bosons  & & & & \\
No fr-in/-out spin 0 & 1.12 to \textbf{0.48} & 3.73 to \textbf{1.61} & 0.73 to \textbf{1.69} & 81 to \textbf{152} \\
No fr-in/-out spin 1 * & 2.24 to \textbf{0.97} & 7.46 to \textbf{3.22} & 0.36 to \textbf{0.84} & 40 to \textbf{76} \\
\hline
\hline
\end{tabular}
}
\end{center}
\end{table}

\section{Adding Dark Matter to the Standard Model}

\begin{figure}
\begin{center}
\scalebox{0.7}
{\includegraphics{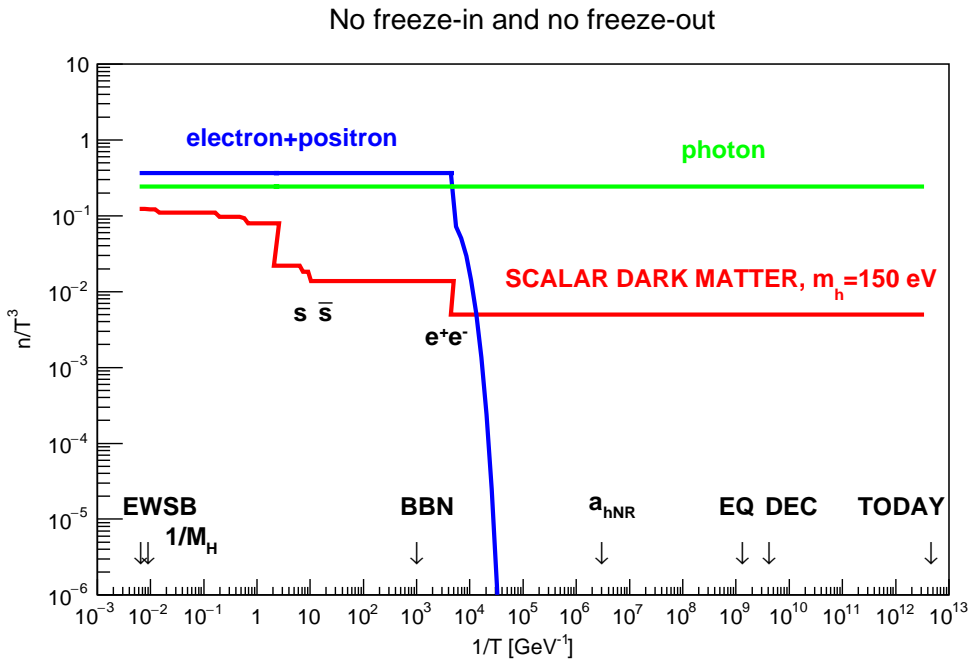}}
\caption{\small{
The no freeze-in and no freeze-out warm dark matter scenario
is illustrated with an example.
$T$ is the photon temperature, and the $n$'s are particle number densities.
}}
\label{example_no_fio}
\end{center}
\end{figure}

The (arguably) simplest renormalizable extensions of the standard model
to add spin 0, $1/2$, or 1 warm dark matter, 
that are in agreement with the no freeze-in and no freeze-out scenario,
are presented in \cite{adding_dm}.
Here we revise the spin 0 case with $m_h = 150 \pm 2$ eV for the
particular scenario developed in \cite{wdm_measurements_and_limits}. The Lagrangian is 
\begin{equation}
\mathcal{L} = \mathcal{L}_\textrm{SM} + \frac{1}{2} \partial_\mu S \cdot S^\mu S - \frac{1}{2} \bar{m}^2_S S^2
- \frac{\lambda_S}{4!} S^4 - \cdots
- \frac{1}{2} \lambda_{hS} \left( \phi^\dagger \phi \right) S^2,
\label{L}
\end{equation}
where $\mathcal{L}_\textrm{SM}$ is the standard model Lagrangian, $S$ is a real scalar
Klein Gordon dark matter field with 
$Z_2$ symmetry $S \leftrightarrow -S$, and 
$\phi$ is the Higgs boson field.
The no freeze-in condition, that
dark matter attains thermal and diffusive equilibrium with the standard model sector 
before the temperature of the universe drops below $M_H$, (arguably) 
requires $|\lambda_{hS}| > 7.4 \times 10^{-7}$ \cite{adding_dm}
(this limit depends on the physics before Electro-Weak Symmetry Breaking (EWSB)).
The condition that the Higgs invisible decay width does not exceed the experimental
bounds requires $|\lambda_{hS}| < 0.03$ \cite{adding_dm}.
The cross-section per unit mass limit $\sigma_\textrm{DM-DM}/m_h < 0.47 \textrm{ cm}^2/\textrm{g}$
\cite{PDG2020} at $a \approx 1$, and equation (13) of \cite{adding_dm},
imply $\lambda_{hS} < 5.7 \times 10^{-5}$, assuming $\lambda_S$ is negligible.
The measured cross-section per unit mass
$\sigma_\textrm{DM-DM}/m_h \approx (1.7 \pm 0.7) \times 10^{-4} \textrm{ cm}^2/\textrm{g}$
\cite{1504.03388} at $a \approx 1$, and equation (13) of \cite{adding_dm},
imply $\lambda_{hS} = (7.8 \pm 0.9) \times 10^{-6}$
(this measurement needs confirmation).
Note that there is a window of opportunity for $\lambda_{hS}$.
The mass of the dark matter particle is
$m_h \equiv M_S = \sqrt{\lambda_{hS} v^2_h/2 + \bar{m}^2_S}$, which
requires fine tuning of $\bar{m}^2_S$ \cite{adding_dm}.
$v_h \approx 246$ GeV is the Higgs boson vacuum expectation value.
See Figure \ref{example_no_fio}.

\section{Conclusions}

Our first measurement of dark matter velocity dispersion, based on
ten spiral galaxy rotation curves measured by the THINGS collaboration,
obtains $v_{h\textrm{rms}}(1) \equiv a_{h\textrm{NR}} c
= 1.25 \pm 0.10 \textrm{ (stat)} \pm 0.75 \textrm{ (syst)}$ km/s,
predicts that this adiabatic invariant is of cosmological origin,
and identifies that this measurement is consistent with the
no freeze-in and no freeze-out warm dark matter scenario \cite{part1}.
Every succeeding study reinforces this view: forty spiral galaxies
of the SPARC sample \cite{adiabatic_invariant}, distributions of galaxy stellar masses 
\cite{fermion_or_boson}, the
formation of first galaxies and reionization \cite{first_galaxies}, and the present
study that includes the distributions of galaxy stellar masses and 
UV luminosities (\ref{vhrms_kfs}), reionization (\ref{kfs_reionization}),
and the velocity dispersion cut-off limit (\ref{vel_disp}).
All of these phenomena are consistent with the no freeze-in and no freeze-out
warm dark matter scenario developed in \cite{wdm_measurements_and_limits}, if
dark matter particles have spin 0, and decouple early on from the standard
model sector.

A summary of measurements is
presented in Table \ref{no_freeze_in_out}.
Note that we have \textit{independently} and redundantly measured three observables of $\Lambda$WDM:
the adiabatic invariant $v_{h\textrm{rms}}(1)$, the delay of structure	formation due to the
free-streaming cut-off factor $\tau^2(k)$, i.e. $k_\textrm{fs}(t_\textrm{eq})$, and the velocity dispersion
cut-off $M_\textrm{vd}$. And the three measured	observables are	consistent with	each other.
However, these measurements are in disagreement
with several limits, of order keV,
on the dark matter ``thermal relic mass" that can be found in
the literature.	These limits on	the ``thermal relic mass" are
really limits on $k_\textrm{fs}(t_\textrm{eq})$ (whether or not we invoke the
no freeze-in and no freeze-out scenario). 
The reason why	the
limits and measurements	differ is that the limits neglect the
non-linear regeneration	of small scale structure
(as studied in \cite{wdm_measurements_and_limits},
and in section \ref{regeneration}). The limits are corrected 
with even a tiny regenerated ``tail"
to $\tau^2(k)$, compared to tails obtained in simulations, 
see first two panels of Figure \ref{n}.
Note that limits may rule out theories, but may not rule out measurements,
if the measurements are correct.
Therefore, may I suggest that the limits be revised, without neglecting the
non-linear regenerated small scale structure (note its huge effect in
\cite{White}), and including the
velocity dispersion cut-off mass (a phenomenon not included in the Press-Schechter formalism).
Let us mention that according to ``The Review of Particle Physics" \cite{PDG2020}, 
limits on dark matter particle mass
are $m_h > 70$ eV for fermions, and $m_h > 10^{-22}$ eV for bosons, and not 
several keV.

The measurements of $v_{h\textrm{rms}}(1)$, or equivalently $k_\textrm{fs}(t_\textrm{eq})$,
determine the dark matter temperature-to-mass ratio, not separately
the temperature or mass. The measured temperature-to-mass ratio happens
to coincide with the no freeze-in and no freeze-out warm dark matter
scenario prediction (as developed in \cite{wdm_measurements_and_limits}) 
if dark matter particles have spin 0,
and decouple early on from the standard model sector. The cases of spin 1/2
and spin 1 are disfavored if nature has chosen the
no freeze-in and no freeze-out scenario of \cite{wdm_measurements_and_limits}, 
see section \ref{no_fifo}.

In summary, a wealth of	measurements redundantly confirm that
dark matter is warm, and, barring a coincidence,
obtain a detailed and precise no freeze-in and no freeze-out
scenario of spin zero warm dark matter particles that decouple
early on from the standard model sector.

\end{document}